\documentclass[12pt,preprint]{aastex}

\usepackage{natbib}
\usepackage{graphicx}
\usepackage{url}

\makeatletter
\newcommand{\Rmnum}[1]{\expandafter\@slowromancap\romannumeral #1@}
\makeatother

\newcommand{\gm}{\textrm{gm}} 
\newcommand{\km}{\textrm{km}} 
\newcommand{\pc}{\textrm{pc}} 
\newcommand{\kpc}{\textrm{kpc}} 
\newcommand{\cmsq}{\ensuremath{\mathrm{cm^2}}} 
\newcommand{\pcubpc}{\ensuremath{\mathrm{pc^{-3}}}} 
\newcommand{\s}{\textrm{s}} 
\newcommand{\ms}{\textrm{ms}} 
\newcommand{\us}{\ensuremath\mathrm{\mu s}} 
\newcommand{\Myr}{\textrm{Myr}} 
\newcommand{\Gyr}{\textrm{Gyr}} 
\newcommand{\ps}{\ensuremath{\mathrm{s^{-1}}}} 
\newcommand{\pssq}{\ensuremath{\mathrm{s^{-2}}}} 
\newcommand{\MHz}{\textrm{MHz}} 
\newcommand{\GHz}{\textrm{GHz}} 
\newcommand{\kelvin}{\textrm{K}} 
\newcommand{\gauss}{\textrm{G}} 
\newcommand{\uJy}{\ensuremath{\mathrm{\mu Jy}}} 
\newcommand{\mJy}{\textrm{mJy}} 
\newcommand{\Jy}{\textrm{Jy}} 
\newcommand{\Msun}{\ensuremath{\mathrm{M_{\Sun}}}} 
\newcommand{\Lsun}{\ensuremath{\mathrm{L_{\Sun}}}} 
\newcommand{\dmu}{\ensuremath{\mathrm{pc\; cm^{-3}}}}
\newcommand{\DM}{\textrm{DM}} 
\newcommand{\sn}{\textrm{S/N}} 
\newcommand{\rmsub}[1]{\ensuremath{_{\mathrm{#1}}}}

\defcitealias{blt+11}{Paper \Rmnum{1}}

\title{A Population of Non-Recycled Pulsars Orginating in Globular
  Clusters}
\shorttitle{Non-Recycled Pulsars Orginating in Globular
  Clusters}

\author{Ryan S.\ Lynch\altaffilmark{1,2}, Duncan R.\
  Lorimer\altaffilmark{3}, Scott M. Ransom\altaffilmark{2}, and Jason
  Boyles\altaffilmark{3}}
\altaffiltext{1}{Department of Physics, McGill University, 3600
  University Street, Montreal, QC, Canada H3A 2T8,
  \texttt{rlynch@physics.mcgill.ca}}
\altaffiltext{2}{National Radio Astronomy Observatory, 520 Edgemont
  Road, Charlottesville, VA 22903}
\altaffiltext{3}{Department of Physics, West Virginia University, 210
  Hodges Hall, Morgantown, WV 26506}

\shortauthors{Lynch et al.}

\keywords{globular clusters: general---pulsars: general}

\begin{document}

\setcounter{footnote}{3}

\begin{abstract}

  We explore the enigmatic population of long-period, apparently
  non-recycled pulsars in globular clusters, building on recent work
  by \citet{blt+11}\ This population is difficult to explain if it
  formed through typical core collapse supernovae, leading many
  authors to invoke electron capture supernovae.  Where
  \citeauthor{blt+11}\ dealt only with non-recycled pulsars in
  clusters, we focus on the pulsars that originated in clusters but
  then escaped into the field of the Galaxy due to the kicks they
  receive at birth.  The magnitude of the kick induced by electron
  capture supernovae is not well known, so we explore various models
  for the kick velocity distribution and size of the population.  The
  most realistic models are those where the kick velocity is $\lesssim
  10\; \km\, \ps$ and where the number of pulsars scales with the
  luminosity of the cluster (as a proxy for cluster mass).  This is in
  good agreement with other estimates of the electron capture
  supernovae kick velocity.  We simulate a number of large-area pulsar
  surveys to determine if a population of pulsars originating in
  clusters could be identified as being separate from normal disk
  pulsars.  We find that the spatial and kinematical properties of the
  population could be used, but only if large numbers of pulsars are
  detected.  In fact, even the most optimistic surveys carried out
  with the future Square Kilometer Array are likely to detect $< 10\%$
  of the total population, so the prospects for identifying these as a
  separate group of pulsars are presently poor.

\end{abstract}

\maketitle

\section{Introduction \label{sec:intro}}

\begin{figure}
\centering
\includegraphics[width=5in]{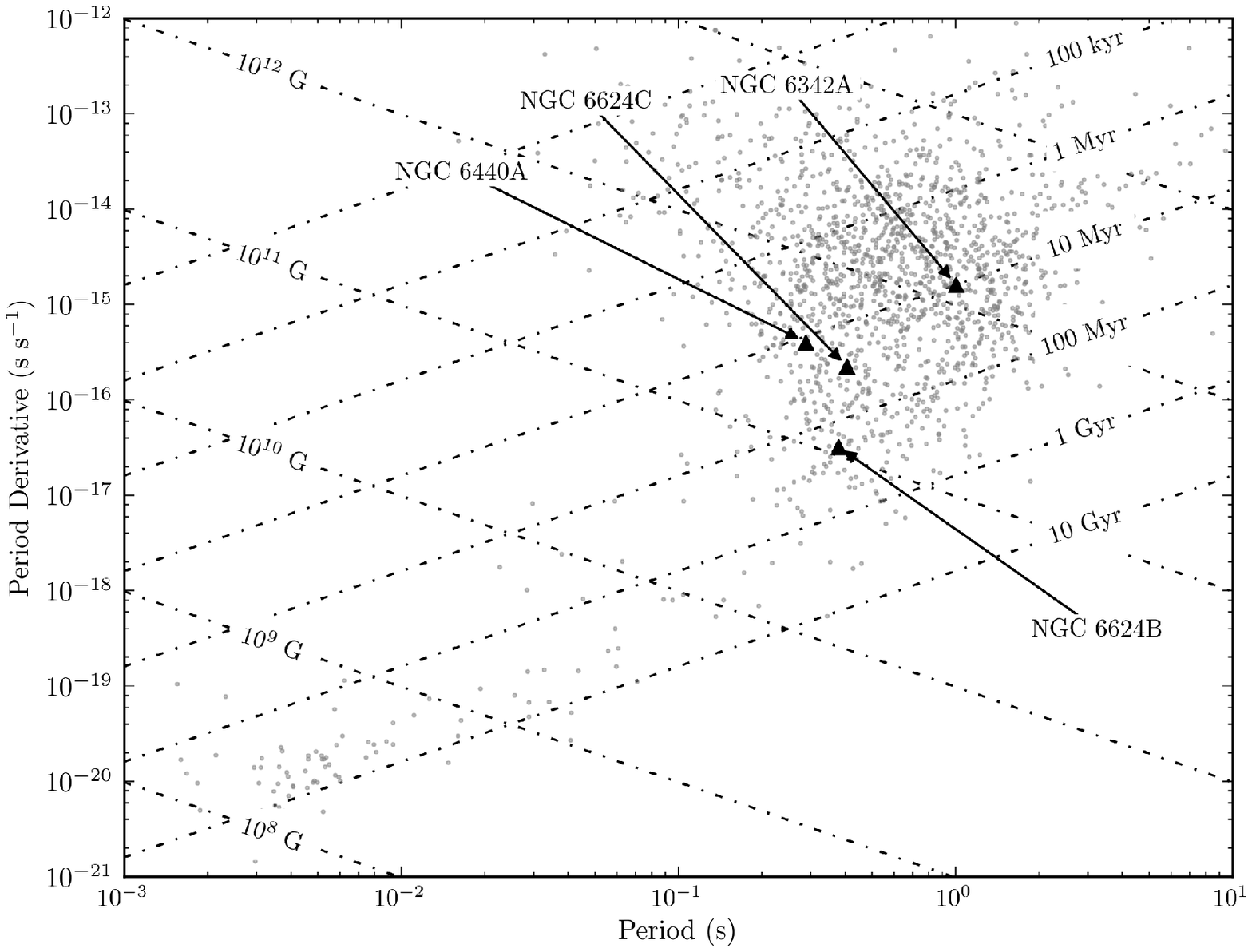}
\caption{The $P$-$\dot{P}$ diagram showing the four NRPs in GCs as
  black triangles.  The gray points are all Galactic pulsars from the
  ATNF Pulsar Database \citep{mht+05}.  GC MSPs have been excluded
  because their $\dot{P}$s are usually contaminated by acceleration in
  the cluster potential.  Lines of constant characteristic age and
  surface magnetic field are also plotted.  \label{fig:ppdot}}
\end{figure}

There are currently 144 pulsars\footnote{For an up-to-date list see
  \url{http://www.naic.edu/~pfreire/GCpsr.html}} known in 28 Galactic
globular clusters (GCs).  The vast majority of these are millisecond
pulsars (MSPs), characterized by short spin periods and small
period-derivatives.  Such a population arises quite naturally in GCs,
which are old stellar systems that are thought to contain a reservoir
of primordial neutron stars (NSs) \citep{hmg+92}.  In the dense
environments in the cores of most GCs, these NSs undergo frequent
exchange interactions, and may then be ``recycled'' into MSPs by
accreting matter from a binary companion \citep{acrs82}.  In addition
to these MSPs, however, there is a small but enigmatic population of
long-period pulsars that seem similar to the ``normal'' pulsars
usually found in the Galactic disk (see Figure \ref{fig:ppdot})
\citep{bbl+94,lmd96,cha03,lfrj12}.  We will refer to these as
\emph{non-recycled pulsars} (NRPs) throughout this paper to
distinguish them from the normal Galactic disk pulsars.  The standard
scenario for forming normal disk pulsars involves the core collapse of
a massive star, giving rise to a young pulsar with a high magnetic
field, which quickly spins down to $P \sim 0.3\; \s$ \citep{f-gk06}.
The inferred lifetimes of normal pulsars are typically $10$ to $100\;
\Myr$ \citep{rl10}, and as such, they are usually associated with the
recent death of massive stars, which themselves have lifetimes
$\lesssim 100\; \Myr$.  This is where the mystery of NRPs in GCs
arises---GCs are composed of old, $\lesssim 1\; \Msun$ stars, and all
stars massive enough to form NSs (along with any pulsars that were
formed) should have died some $\sim 10\; \Gyr$ ago.  The fact that
several NRPs \emph{are} observed in GCs requires an alternative to the
standard core collapse model.

Several authors \citep{lmd96,ihr+08} have invoked the collapse of a
massive O-Ne-Mg white dwarf (WD) via electron capture (so called
electron capture supernovae, or ECS \citep{n84,n87}).  Recent
theoretical work supports the notion that ECS are essential for
understanding the full population of neutron stars in GCs
\citep{ihr+08}.  Unfortunately, ECS have never been directly observed
and their energetics remain uncertain, although there is good reason
to believe that they are about an order of magnitude less energetic
than core collapse supernovae (CCS) \citep{kjh06,dbo+06}.  This
probably leads to small natal kicks when combined with a faster and
more symmetric explosion than in CCS \citep{plp+04}.  NRPs may be an
important observational constraint on the properties of ECS.

Motivated by the large number of recent and sensitive pulsar searches
of GCs, \citet[hereafter \citetalias{blt+11}]{blt+11} used statistical
techniques to explore the underlying population of NRPs in GCs.  The
results are based solely on observations and some simplifying
assumptions about the luminosity function and lifetime of NRPs, and as
such provide a constraint on the birthrate of NRPs in GCs, regardless
of how they are formed.  One of the key results from this study was
the compilation of probability distributions for the birthrates of
NRPs in the majority of GCs.

\citetalias{blt+11} gave detailed consideration only to those pulsars
that gained a sufficiently small natal kick as to be retained by their
host clusters.  Given the high upper limits on the birthrate of GC
NRPs and the relatively shallow potentials of most GCs, there is an
intriguing possibility that a large population of NRPs may have
escaped from their progenitor clusters at birth and entered the field
of the Galaxy, where they could contribute to the observed population
of normal disk pulsars.  Building on the results of
\citetalias{blt+11}, we have carried out Monte Carlo simulations to
explore the properties of this purported population, and the
feasibility of detecting it as a separate group.  In
\S\ref{sec:paper1} we provide a brief overview of the techniques and
results from \citetalias{blt+11}.  In \S\ref{sec:sims} we describe our
simulations and present the results in \S\ref{sec:results}.  We
discuss the implications of these results in \S\ref{sec:discussion}
and summarize in \S\ref{sec:conc}.

\section{Overview of \citetalias{blt+11} \label{sec:paper1}}

The goal of \citetalias{blt+11} was to characterize the underlying
population of NRPs in a cluster using observational constraints.
Obviously, this approach is only valid for clusters that have been
searched for pulsars.  \citetalias{blt+11} compiled results from
searches of 97 clusters (out of 156 listed in \citet{har10}) carried
out by numerous groups (see \citetalias{blt+11} for a complete list
and references).  For each search, the detection probability was
defined as
\begin{eqnarray}
\theta(L\rmsub{min},f(L)) = \frac{\int_{L\rmsub{min}}^{\infty}{f(L)\: dL}}
                                 {\int_0^{\infty}{f(L)\: dL}},
\end{eqnarray}
where $L\rmsub{min}$ is the limiting luminosity of the search, and
$f(L)$ is the luminosity distribution of NRPs.  In words, $\theta$ is
simply the fraction of pulsars that lie above $L\rmsub{min}$.  Five
log-normal distributions with different parameters were used for
$f(L)$; one given by \citet{f-gk06} and four by \citet{rl10}.  This
gave rise to a range of values for $\theta$.

Bayes' theorem was then used to obtain a probability density function
(PDF) for the total number of \emph{potentially observable} NRPs
($N$), given the number of actual detections ($n$), and $\theta$.
Potentially observable means all NRPs that currently reside in
clusters and whose radio beams cross our line of sight (i.e., all
those that could be observed with infinite sensitivity).  We shall
discuss corrections to this number in \S\ref{sec:sims}.  This PDF is
\begin{eqnarray}
  \mathcal{P}(N,\theta|n) \propto \mathcal{L}(n,N|\theta) 
                                  \mathcal{P}(N,\theta),
\end{eqnarray}
where $\mathcal{L}(n,N|\theta)$ is the likelihood of detecting $n$
pulsars from a population of $N$ given $\theta$, and
$\mathcal{P}(N,\theta)$ is the joint prior PDF for $N$ and $\theta$.
Since pulsar searches have only two possible outcomes (success or
failure), \citetalias{blt+11} chose the binomial distribution for the
likelihood:
\begin{eqnarray}
  \mathcal{L}(n,N|\theta) = \frac{N!}{n!(N-n)!} \theta^n (1-\theta)^{N-n}.
\end{eqnarray}
To characterize the joint prior, $\mathcal{P}(N,\theta)$,
\citetalias{blt+11} first assumed that $N$ and $\theta$ were
independent.  The prior for $N$ was taken to be a uniform distribution
on $[n,\infty)$, so that the only restriction is $N \geq n$.  The
prior for $\theta$ was also taken to be uniformly distributed on
$[\theta\rmsub{min},\theta\rmsub{max}]$, where $\theta\rmsub{min}$ and
$\theta\rmsub{max}$ correspond to the lowest and highest values of
$\theta$ for different choices of $f(L)$.  Having obtained
$\mathcal{P}(N,\theta|n)$, \citetalias{blt+11} then marginalized over
$\theta$ to obtain a final PDF for $N$ (denoted as $\mathcal{P}(N|n)$)
for each of the 97 clusters in the sample.

\section{Simulating NRPs that Escape from Clusters \label{sec:sims}}

\begin{figure}
\centering
\includegraphics[height=7.0in]{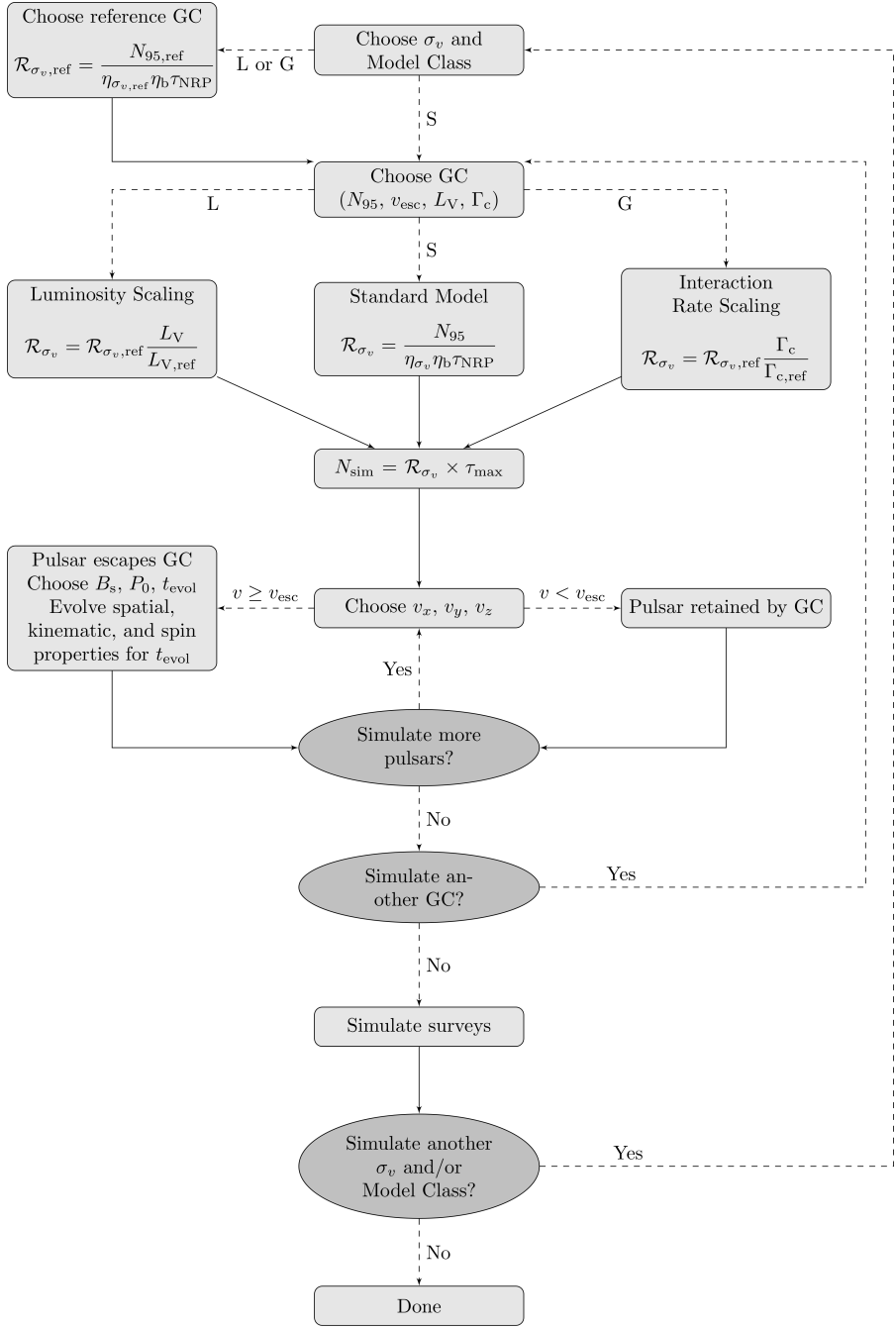}
\caption{A schematic diagram describing our simulation strategy.  The
  abbreviations for model classes are: S\,=\,standard models (no scaling);
  L\,=\,luminosity scaling; G\,=\,core interaction rate scaling.
  \label{fig:chart}}
\end{figure}

As mentioned in \S\ref{sec:paper1}, it is necessary to correct
$\mathcal{P}(N|n)$ to account for the fraction of NRPs whose radio
beams do not cross our line of sight, and for those who are no longer
bound to their progenitor GCs.  This second correction is necessary
because GCs have fairly shallow potentials, with escape velocities
$v\rmsub{esc} \sim 30\; \km\, \ps$.  Some NRPs will inevitably escape
the cluster due to the natal kicks they receive during formation and
enter the field of the Galaxy.  This is precisely the population that
we are interested in here.  We have used Monte Carlo simulations to
model the evolution of NRPs that escaped from GCs and to determine if
this population could be distinguished observationally from other
pulsars in the Galaxy.  We proceed in four steps: \textit{i}) we
determine the number of NRPs that will enter the Galaxy; \textit{ii})
we model the spatial and kinematical evolution of this population;
\textit{iii}) we model the rotational and electromagnetic evolution;
and \textit{iv}) we ``detect'' this evolved population by simulating a
number of large-area pulsar surveys.  As described in the following
sections, we actually explore several different types of models for
the population, and for each model we also explore several natal kick
distributions.  To ensure robust statistical results, each combination
of model class and kick distribution is simulated ten times, and the
final results are taken to be the mean of these runs.  Figure
\ref{fig:chart} is a schematic overview of our simulations.  We
discuss each step below.

\subsection{Determining the Number of Escaped NRPs \label{sec:Nesc}}

We choose to characterize $\mathcal{P}(N|n)$ by the upper bound of the
95\% confidence interval, which we denote as $N_{95}$ (note that the
birth rates discussed in \citetalias{blt+11} used the \emph{median} of
this distribution).  This is corrected for beaming and escaping
pulsars, and divided by the mean lifetime of NRPs to obtain a birth
rate for each cluster that we consider,
\begin{eqnarray}
  \mathcal{R}_{\sigma_v} = \frac{N_{95}}
                                {\eta\rmsub{beam} \eta_{\sigma_v} 
                                 \tau\rmsub{NRP}},
\end{eqnarray}
where $\mathcal{R}_{\sigma_v}$ is the birth rate, $\eta\rmsub{beam} =
0.1$ is the beaming fraction \citep{tm98}, $\eta_{\sigma_v}$ is the
correction for escaping pulsars, and $\tau\rmsub{NRP}$ is the average
pulsar lifetime.  We derive an average lifetime of $43\; \Myr$ by
taking the total number of radio-loud pulsars ($1.2 \times 10^6$) and
dividing it by the Galactic pulsar birthrate ($2.8\;
\mathrm{century}^{-1}$) \citep{f-gk06}.  The subscript $\sigma_v$
refers to the mean birth velocity dispersion.  The actual value of
$\sigma_v$ for NRPs is unknown, since they are probably not formed via
typical CCSNe , so we vary this parameter in our simulations.  As we
shall see, there is good reason to believe $\sigma_v \sim 10\; \km\,
\ps$.  We assume that birth velocities can be described by a
Maxwell-Boltzmann distribution so that the fraction of NRPs with $v
\leq v\rmsub{esc}$ is
\begin{eqnarray}
\eta_{\sigma_v} & = & \frac{\int_0^{v\rmsub{esc}}{f\rmsub{MB}(v)\: dv}}
                                   {\int_0^{\infty}{f\rmsub{MB}(v)\: dv}} \nonumber \\
\,              & = & \mathrm{erf}\left ( \frac{v\rmsub{esc}}
                                               {\sqrt{2 \sigma_v}} \right )
                      - \sqrt{\frac{2}{\pi}}\frac{v\rmsub{esc}}{\sigma_v}\: 
                        \exp{\left (\frac{-v\rmsub{esc}^2}{2 \sigma_v^2} \right ) },
\end{eqnarray}
where $f\rmsub{MB}(v)$ is the Maxwell-Boltzmann distribution and
$\mathrm{erf}$ signifies the error function
\begin{eqnarray}
\mathrm{erf}(x) = \frac{2}{\sqrt{\pi}} \int_0^x e^{-t^2}\: dt
\end{eqnarray}
Escape velocities for each cluster were taken from \citet{gzp+02}.
For clusters without a reported value, we assumed $v\rmsub{esc} = 30\;
\km\, \ps$, which is roughly the median value from \citet{gzp+02}.

Having obtained $\mathcal{R}_{\sigma_v}$, all that remains is to
choose a time scale, $\tau\rmsub{max}$, over which we will consider
the evolution of this population.  The number of pulsars to simulate
per cluster then becomes $N\rmsub{sim} = \mathcal{R}_{\sigma_v} \times
\tau\rmsub{max}$.  In our simulations, we choose $\tau\rmsub{max} =
200\; \Myr$.  As we explain in \S\ref{sec:pevol}, many pulsars will
cease radio emission on timescales shorter than this.  We choose a
large value for $\tau\rmsub{max}$ to ensure that we treat long-lived
NRPs properly.  

We refer to the above approach as our ``standard model'' (hereafter
Sa).  $N\rmsub{sim}$ is obtained directly from the results of
\citetalias{blt+11}, which is to say, without using any information
about the clusters except for their escape velocities.  However, the
value of $N_{95}$, and hence $N\rmsub{sim}$, depends heavily on
$L\rmsub{min}$, in the sense that $N_{95}$ will be larger (and not
very constraining) for shallowly searched GCs.  Therefore, we included
three refinements to Sa in our study.

The first we call the ``modified standard model'' (Sb).  For this, we
simply exclude 16 clusters with $L\rmsub{1.4\; GHz,min} \geq 7.5\;
\mJy\, \kpc^2$, which reduces the number of shallowly searched
clusters with very high values of $N_{95}$.  While this sensitivity
cut-off is somewhat arbitrary, it strikes a good balance between
keeping enough clusters for the simulations to have meaningful
results, while preventing unwieldy computations due to extremely large
numbers of simulated escaping NRPs.

In the next two models, we choose a reference cluster, and scale all
other values of $N\rmsub{sim}$ by some chosen parameter.  We use M22
as a reference cluster because it has the lowest value of
$L\rmsub{min}$ and consequently the most tightly constrained
$N\rmsub{sim}$ (relevant parameters of M22 can be found in Table
\ref{table:M22}).  In the first approach, we scale by the V-band
luminosity of each cluster, $L\rmsub{V}$, so that
\begin{eqnarray}
N^{\mathrm{L}}\rmsub{sim} = N\rmsub{sim, M22} \frac{L\rmsub{V}}
                                                   {L\rmsub{V,M22}}.
\end{eqnarray}
This is the ``luminosity scaling model'' (L).  Our reasoning is that
at the very least, the number of NRPs produced in a cluster should
scale with the number of progenitors in the cluster.  However, to
determine the number of progenitors would require first identifying
those progenitors, and then making some assumptions about how their
number scales with other properties of the cluster.  We sought a more
general approach, hence our decision to use the luminosity of the
cluster---brighter clusters should be more massive (to within a factor
of the cluster's mass-to-light ratio), and more massive clusters
should contain more NRP progenitors.  Furthermore, the V-band
luminosity is fairly easily determined as long as the distance to the
cluster and reddening effects are well constrained\footnote{The values
  we use are from \citet{har10} and have been corrected for
  reddening.}, and is also readily available for nearly all Milky Way
GCs.

Finally, we also scale by the so-called core interaction rate,
$\Gamma\rmsub{c} \propto \rho\rmsub{c}^{1.5} r\rmsub{c}^2$, where
$\rho\rmsub{c}$ is the central density of the cluster and $r\rmsub{c}$
is the core radius; thus we have
\begin{eqnarray}
N^{\mathrm{G}}\rmsub{sim} = N\rmsub{sim, M22} \frac{\Gamma\rmsub{c}}
                                                   {\Gamma\rmsub{c,M22}}.
\end{eqnarray}
We call these the ``interaction rate scaling models'' (G).
$\Gamma\rmsub{c}$ has been shown to correlate well with the number of
low-mass X-ray binaries and MSPs in a cluster \citep{p+03,a+10},
further supporting the notion that both populations are related to
binary exchange interactions.  A leading explanation for NRPs in GCs
are ECS, particularly via accretion or merger induced collapse.  These
scenarios also require mass-transfer or merger in a binary system.
Hence, it is interesting to explore models in which NRPs also have
some dependence on $\Gamma\rmsub{c}$.

For each of these models, relevant parameters ($N95$, $L\rmsub{V}$,
and $\Gamma\rmsub{c}$) were only available for a subset of all 156
Milky Way GCs.  It will be necessary to adjust our final results to
account for unmodeled clusters, but we save this step for
\S\ref{sec:results}.  Table \ref{table:models} summarizes each class
of models.

\subsection{Spatial and Kinematical Evolution \label{sec:xvevol}}

In this step, we model the evolution of the escaping NRPs as they
travel through the Galaxy.  We begin by defining a Galactocentric
coordinate system with
\begin{eqnarray}
x & = & D \cos{b} \sin{\ell},                         \\
y & = & D_{\Sun} - D \cos{b} \cos{\ell},\mathrm{~and} \\
z & = & D \sin{b}
\end{eqnarray}
where $D$ is the distance, $D_{\Sun} = 8.5\; \kpc$ is the Sun's
distance from the Galactic center, $b$ is Galactic latitude, and
$\ell$ is Galactic longitude.  Note that this differs from the typical
Cartesian Galactic coordinates, where it is usually the $x$-axis that
connects the Sun and Galactic center.  This was done for compatibility
with previously developed software tools.

We assign each NRP initial coordinates that are equal to the
coordinates of the host GC.  Initial 3-D velocity components
($v_{x,\mathrm{i}}$, $v_{y,\mathrm{i}}$, $v_{z,\mathrm{i}}$) are
chosen at random from a normal distribution with a zero mean and
standard deviation of $\sigma_v$ (this is equivalent to a
Maxwell-Boltzmann distribution for the full space velocity).  If the
total space velocity with respect to the GC, $v\rmsub{i} =
(v_{x,\mathrm{i}}^2 + v_{y,\mathrm{i}}^2 + v_{z,\mathrm{i}}^2)^{1/2}
\geq v\rmsub{esc}$, then we consider the pulsar to have escaped from
its host cluster and entered the field of the Galaxy.  Although we
keep track of the number of pulsars retained by their host clusters,
we do not consider them further.  All subsequent analysis on the
number of emiting and detectable NRPs deals only with those that
escape.  We treat each GC as being fixed at its current position in
the Galaxy.  In reality, the clusters are on their own orbits, and the
pulsars will inherit the systemic velocity of their progenitor
clusters.  Each pulsars enters the field of the Galaxy with
$v\rmsub{field} = (v\rmsub{i}^2 - v\rmsub{esc}^2)^{1/2}$.

Having chosen initial spatial and velocity components, we integrate
the motion of each pulsar through the Galactic potential.  Following
\citet{ci87} and \citet{kg89} \citep[see also][]{wkk00}, we use
a 3-component Galactic potential of the form
\begin{eqnarray}
\phi^i(R,Z) = - G M^i \left [ \left (a^i + 
                              \sum\limits_{j=1}^3 \left [ \beta_j^i 
                                                          \sqrt{Z^2 + h_j^i}\, \right ]
                              \right )^2 + b^i + R^2 \right ]^{-1/2},
\end{eqnarray}
where the superscript $i = \mathrm{D, B, N}$ indicates the
contribution from the disk-halo, bulge, or nucleus, respectively.  The
full potential is the sum of these three components.  The values of
$M^i$, $a^i$, $b^i$, $\beta^i_j$, and $h^i_j$ can be found in Table
\ref{table:phi_params}.  The integration time, $t\rmsub{evol}$, is
chosen at random from a uniform distribution on the interval $[0,200\;
\Myr]$.  In other words, we assume that an NRP may be born at any
point in the past $200\; \Myr$, and evolve it forward to the present.
This also assumes that the birthrate of NRPs is constant over this
interval.  The final spatial and kinematical properties are then
recorded for later analysis.

\subsection{Rotational Evolution and Energetics \label{sec:pevol}}

Birth spin periods and surface magnetic fields ($B\rmsub{s}$) are
chosen at random for each escaping NRP.  Following \citet{f-gk06}, we
use a normal distribution for $P_0$ with a mean of $0.3\; \s$ and
standard deviation of $0.15\; \s$.  Negative periods are rejected and
redrawn.  We also use a log-normal distribution for $B\rmsub{s}$, with
a mean in the base-10 logarithm of $12.65$ and standard deviation
$0.55$.

We evolve the pulsar's rotation under the assumption of pure magnetic
dipole braking and a constant magnetic field, so that the observed
period and period derivative are
\begin{eqnarray}
  P       & = & \left (\frac{P_0^2 + 16 \pi B\rmsub{s}^2 R^6 t\rmsub{evol}}
    {3 c^3 I} \right)^{1/2} \mathrm{~and} \\
  \dot{P} & = & \frac{P^2 - P_0^2}{2 P t\rmsub{evol}},
\end{eqnarray}
where $R = 10\; \km$ and $I = 10^{45}\; \gm\, \cmsq$ are the assumed
radius and moment of inertia for the pulsar, respectively, and
$t\rmsub{evol}$ is defined as above.  Hence, we arrive at the final
$P$ and $\dot{P}$ at the end of our simulation.

We calculate the observed luminosity using a power-law model that
depends on $P$ and $\dot{P}$.  We prefer this model because it relates
the luminosity to the rotational energy loss of the pulsar.  Once
again, we turn to \citet{f-gk06} for the exact form of this power-law:
\begin{eqnarray}
  L = 10^{L\rmsub{cor}} L_0 P^{-3/2} \left ( \frac{\dot{P}}
                                              {10^{-15}\; \s\, \ps} \right ) ^{1/2}.
\end{eqnarray}
$L_0 = 0.18\; \mJy\, \kpc^2$ is a ``standard'' luminosity and
$L\rmsub{cor}$ is a correction factor that accounts for uncertainty in
the model.  It is drawn at random from a normal distribution with zero
mean and standard deviation $\sigma_{L\rmsub{cor}} = 0.8$.

Pulsars have finite lifetimes, and at some point will cease radio
emission.  This seems to correspond to a ``death-line'' in the
$P$-$\dot{P}$ diagram, which is well described theoretically
\citep{bwhv92} and empirically as
\begin{eqnarray}
  \frac{B\rmsub{s}}{P^2} = 1.7 \times 10^{11}\; \gauss\, \pssq.
\end{eqnarray}
Any pulsars with $B\rmsub{s} P^{-2}$ less than this value will no
longer be visible.  We track the evolution of these pulsars, but do
not include them for consideration in the next step.

\subsection{Simulated Surveys \label{sec:surveys}}

The final step is to ``search'' for potentially visible NRPs by
simulating various large-area surveys.  A pulsar can be detected only
if its signal-to-noise ratio (\sn) lies above the minimum \sn\
threshold of a given survey.  Following \citet{lk04},
\begin{eqnarray}
\sn = S_{\nu} \frac{G \sqrt{N\rmsub{pol} \Delta \nu t\rmsub{int}}}
                   {\beta (T\rmsub{sys} + T\rmsub{sky})}
      \sqrt{\frac{P}{W} - 1}
\end{eqnarray}
where $S_{\nu}$ is the flux density of the pulsar, $G$ is the
telescope gain, $N\rmsub{pol}$ is the number of summed polarizations,
$\Delta \nu$ is the bandwidth, $t\rmsub{int}$ is the integration time,
$\beta$ is a factor that accounts for quantization losses,
$T\rmsub{sys}$ and $T\rmsub{sky}$ are the system and sky temperatures,
respectively, and $W$ is the observed pulse width.  Most of these
factors are intrinsic to the specific survey in question, but
$T\rmsub{sky}$ and $W$ depend on the position and properties of the
pulsar.  Sky temperatures at the position of each pulsar are taken
from the $408\; \MHz$ survey of \citet{hssw82} and scaled to the
appropriate frequency assuming a power-law with a spectral index of
$-2.6$.  The pulse width is described as the quadrature sum of 
\begin{eqnarray}
  W^2 = W\rmsub{int}^2 + t\rmsub{samp}^2 + t\rmsub{DM}^2 + \tau\rmsub{s}^2,
\end{eqnarray}
where $W\rmsub{int}$ is the intrinsic pulse width, $t\rmsub{samp}$ is
the instrumental sampling time, $t\rmsub{DM}$ is the dispersive
smearing within a given frequency channel, and $\tau\rmsub{s}$ is the
scattering time.  We model the intrinsic width as
\begin{eqnarray}
\log{W\rmsub{int}} = \log{\left [ 0.06 \left (\frac{P}{\ms} \right)^{0.9} \right ]} +
                     \varsigma,
\end{eqnarray}
where $\varsigma$ is a random variable drawn from a normal
distribution with a standard deviation of $0.3$ \citep{l+06}.  The
sampling time for each survey is listed in Table \ref{table:surveys}.
The dispersion measure (DM) is calculated using the known distance to
the pulsar and the NE2001 model of Galactic free electron density
\citep{cl02}.  The dispersive smearing is then simply
\begin{eqnarray}
t\rmsub{DM} \simeq 8.3 \times 10^3\; \s 
                   \left (\frac{\DM}{\dmu} \right ) 
                   \left (\frac{\Delta \nu\rmsub{chan}}{\MHz} \right ) 
                   \left (\frac{\nu}{\MHz} \right )^{-3},
\end{eqnarray}
where $\Delta \nu\rmsub{chan}$ is the channel width and $\nu$ is the
center frequency \citep{lk04}.  Various empirical relations between
$\tau\rmsub{s}$ and DM appear in the literature.  We use the
relationship given by \citet{cor02}:
\begin{eqnarray}
  \lefteqn{\log{\left (\frac{\tau\rmsub{s}}{\us} \right )} =} \nonumber \\
    && -3.59 + 0.129 \log{\left (\frac{\DM}{\dmu} \right )}
    + 1.02 \left [ \log{\left (\frac{\DM}{\dmu} \right )} \right ]^2
    - 4.4 \log{\left (\frac{\nu}{\GHz} \right )}.
\end{eqnarray}

One factor that we do not model here is the affect of radio frequency
interference (RFI), which can be particularly problematic for blind
searches of long-period pulsars.  Obviously, some surveys will be more
affected by RFI because of proximity to man-made sources and/or poor
instrumental resistance to strong RFI.  Nonetheless, careful data
analysis can help to mitigate these effects.

After determining the relevant parameters for each NRP, we use the
\texttt{survey} tool from the \texttt{psrpop} software
package\footnote{\url{http://psrpop.sourceforge.net/}} \citep{l+06} to
simulate the surveys.  Pulsar luminosities are scaled to the
appropriate frequency assuming a power-law spectral index of $-1.6$.

We simulate five surveys: the Parkes Multibeam Survey (PMSURV)
\citep{l+06}; the P-ALFA survey \citep{c+06}, the GBT North Celestial
Cap Survey (GBNCC); a visible-sky GBT survey similar to the GBNCC
survey (GBTALL); and a hypothetical all-sky survey using a future
Square Kilometer Array (SKA) \citep{sks+09}.  The characteristics of
each survey can be found in Table \ref{table:surveys}.  Each survey is
simulated 100 times, for each run of the simulation, allowing us to
characterize the median number of detected pulsars.  Because the
spatial, kinematical, and rotational evolution of the NRPs is
simulated ten times for each combination of model class and $\sigma_v$
(see above), we have ten values for the median number of detected
pulsars from each combination.  The final reported number of detected
NRPs ($N\rmsub{det}$) is the mean of these ten values plus the
standard error, and represents an upper limit.

\section{Results \label{sec:results}}

The results of our simulations can be found in Table
\ref{table:results}.  The values reported here have been multiplied by
a scale factor ($156/N\rmsub{GC}$) that accounts for clusters not
included in that model class.  This is equivalent to assuming that
un-modeled clusters follow the average results for that model class.
The total number of \emph{all} NRPs that escape from their progenitor
clusters ($N\rmsub{esc}$) and that are retained by their host clusters
($N\rmsub{ret}$) are listed in the first two columns.  These include
pulsars whose radio beams do not cross our line of sight as well as
those that have crossed the death line; these will be dealt with in
\S\ref{sec:surv_results}.  Hence, these numbers represent 95\%
confidence interval upper limits on the total number of NRPs formed in
the last $200\; \Myr$\footnote{$N\rmsub{esc}$ and $N\rmsub{ret}$ scale
  linearly with the choice of $\tau\rmsub{max}$, so it is trivial to
  adjust these when considering a different timescale.}.

It is immediately obvious that, for a given model class,
$N\rmsub{esc}$ increases with increasing $\sigma_v$.  This is not at
all surprising, since a larger kick velocity dispersion will lead to
more high velocity pulsars that can escape the cluster potential.  In
contrast, $N\rmsub{ret}$ remains nearly constant with $\sigma_v$.
This is because we have essentially normalized all values by the
``observable'' (i.e., based on observations) value of $N_{95}$, which
characterizes the number of retained NRPs.  We point out that
$\mathcal{R}_{\sigma_v}$ also increases with $\sigma_v$, so that there
are more pulsars in total simulated for higher velocity dispersions.
Model class Sa gives rise to the largest total number of NRPs by far.
This is entirely due to the very high values of $N_{95}$ for most GCs.
Even when we exclude the most unconstraining clusters from
consideration in model class Sb, the total number of NRPs is still
very large.  Model class G produces fewer pulsars than Sb, while model
class L produces the fewest of all.  As we shall see in
\S\ref{sec:real_models}, this has important consequences for the
viability of ECS as an explanation for NRPs.  For now, we turn our
attention to exploring the characteristics of the escaped population
in more detail.

\subsection{Spatial and Kinematical Properties
  \label{sec:space_kin_props}}

\begin{figure}
\centering
\includegraphics[width=3.0in]{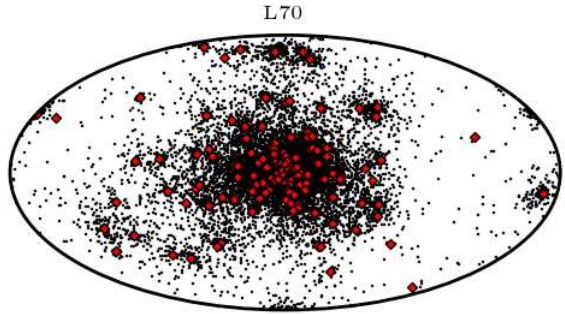}
\caption{The position of all escaped pulsars in Galactic coordinates
  (as viewed from Earth) for model L70.  Other models have a similar
  spatial distribution of pulsars but with varying numbers.  The
  Galactic center is at the center of the plot.  Globular clusters are
  represented by red diamonds, while pulsars are shown as black dots.
  \label{fig:pos_L}}
\end{figure}

Figure \ref{fig:pos_L} shows a sample sky distribution of all escaping
NRPs for model L70 in Galactic coordinates (other models have very
similar forms with different numbers of pulsars).  The pulsars tend to
group around their progenitor clusters.  In models with higher
$\sigma_v$ the pulsars travel further from their host clusters, as
expected, but there are also more pulsars in total.

We also model calculate the proper motion of each pulsar.  The proper
motions are a combination of the kick velocity of the pulsar (evolved
as the pulsar travles through the Galaxy) and the systemic velocity of
the GC itself (both projected onto the plane of the sky).  Pulsars
from a common progenitor cluster will thus have similar systemic
velocity components.  The velocities of many GCs arepdf observed to be
$\sim 100\; \km\, \ps$, larger than the kick velocities received by
the NRPs at birth.

\subsection{Number of Detectable NRPs in Surveys
  \label{sec:surv_results}}

The results of our simulated surveys can be found in Table
\ref{table:results}.  It is at this stage that we account for pulsars
that have crossed the death line.  The third column gives the number
of all radio emitting pulsars ($N\rmsub{emit}$).  The number of
\emph{detected} pulsars for various surveys is given in the following
columns\footnote{$N\rmsub{emit}$ and $N\rmsub{det}$ are not very
  sensitive to the choice of $\tau\rmsub{max}$ unless it is less than
  the lifetime of a typical NRP.}.  These numbers have already been
multiplied by a constant $\eta\rmsub{beam} = 0.1$ to account for
pulsars whose radio beams do not cross our line of sight.  It is
immediately obvious that significant numbers of GC NRPs are detected
only when $N\rmsub{emit}$ is large (corresponding to large
$\sigma_v$).  The most successful survey is, not surprisingly, our
hypothetical all-sky SKA survey.  Such a survey would presumably be
designed to detect nearly all potentially observable disk pulsars in
the Galaxy, but only $\sim 8\%$ of emitting GC NRPs are detected.  The
next most promising survey is the hypothetical GBTALL, but this
detects only $\sim 0.03\%$ of emitting NRPs.  The success of the
PMSURV and P-ALFA are less than but comparable to the GBTALL.  The
GBNCC detects very few, if any, NRPs.  We can attribute this to the
design of the survey, whose goal it is to find nearby and bright MSPs;
the $1374\; \MHz$ limiting flux density is an order of magnitude
higher than PMSURV, P-ALFA, or the SKA survey.  However, the GBTALL
has an identical set-up and does much better.  We attribute this to
the large survey area and better coverage at low declinations, where
most GCs are found.  The PMSURV and P-ALFA both utilize high observing
frequencies and long integration times to search for highly dispersed,
and thus generally more distant pulsars.  At first glance it may seem
strange that so few NRPs are detected, given that we already know of
four that reside in GCs.  These pulsars were detected in targeted
surveys with very long integration times.  Furthermore, for low
$\sigma_v$, we predict $N\rmsub{esc} < N\rmsub{ret}$; in models where
$N\rmsub{esc} \approx N\rmsub{ret}$, the number of detected NRPs is
closer to what is actually observed in clusters.

We can conclude that, to have any hope of detecting large numbers of
escaped NRPs, surveys must be very sensitive and cover as much sky as
possible.  Even then, success will depend strongly on the size of the
underlying population, which depends on the kick velocity of NRPs.

\section{Discussion \label{sec:discussion}}

\subsection{Which Models are Realistic? \label{sec:real_models}}

One of the primary findings of \citetalias{blt+11} was that very high
values of $\sigma_v$ greatly overproduce NRPs in GCs according to any
reasonable metric.  For example, they find a birth rate of $422$ and
$0.25\; \mathrm{psr\, century^{-1}}$ for $\sigma_v = 130$ and $10\;
\km\, \ps$, respectively (keep in mind that these are for the median
number of NRPs, whereas we use the higher 95\% confidence upper
limit).  For comparison, the birth rate of normal disk pulsars is
$2.8\; \mathrm{psr\, century^{-1}}$ \citep{f-gk06}.  More relevant for
the case of NRPs is how these results compare with the population of
WDs that are the likely progenitors to NRPs via ECS.
\citetalias{blt+11} estimated the total number of WDs that could
potentially form NSs via ECS\footnote{Assumed to be all WDs with $1.0
  \leq M \leq 1.4\; \Msun$.} to be $\sim 2.1 \times 10^5$ across the
entire GC population.  If we assume that all of these WDs eventually
become NSs and that the birthrate of NRPs is constant, then the
timescale for exhausting this population is simply $2.4 \times 10^5/
\mathcal{R}_{\sigma_v}$.  These timescales, $\tau\rmsub{exhaust}$, are
given for each of our models in Table \ref{table:birthrates}.  Even
for the smallest $\mathcal{R}_{\sigma_v}$ in our standard models,
$\tau\rmsub{exhaust} \lesssim 100\; \Myr$.  This is only $\sim 1\%$
the age of a typical GC ($\sim 10\; \Gyr$).  Although these birth
rates are only upper limits, such a large discrepancy seems difficult
to explain.  Recall, however, that our standard models are heavily
affected by the limiting luminosity of cluster searches, so we believe
that these results simply indicate that the standard models are not
very well constrained.  Model class G, where we scale the number of
NRPs formed in M22 by the core interaction rate of a cluster, produces
fewer NRPs than our standard models.  Nonetheless, inspection of Table
\ref{table:birthrates} indicates that the implied birthrates are still
probably too high.

However, the situation improves for model class L, where we scale by
the luminosity of a cluster.  High velocity dispersion cases still
seem to overproduce the number of NRPs considerably, but for $\sigma_v
< 30\; \km\, \ps$ the implied birth rates are low enough that the
resevoir of WDs may last for $> 15\%$ of the age of the typical
cluster.  When one considers that the birth rates reported here are
only 95\% confidence upper limits, it seems that ECS are a viable
explanation for NRPs after all.  Given that the true efficiency in
going from WDs to NRPs is probably $< 100\%$, we favor a typical ECS
kick velocity of $\lesssim 10\; \km\, \ps$.  Other authors have
inferred small $\sigma_v$ for ECS as well.  \citet{prp+02} studied the
eccentricities of high-mass X-ray binaries that are believed to form
via ECS and concluded that $\sigma_v \lesssim 50\; \km\, \ps$.  More
recently, \citet{mtp09} used a similar line of reasoning to argue for
$\sigma_v \sim 15\; \km\, \ps$ (though the authors note that this may
be in conflict with the observed misalignment between the rotational
and orbital spin axes in these systems).  PSR J0737$-$3039B is also
believed to have formed via ECS \citep{pdl+05} and \citet{wwk10} find
a 95\% confidence interval of $5$--$120\; \km\, \ps$ for the kick of
this neutron star.  Our results are in general agreement with the
notion that ECS must be less energetic and lead to smaller neutron
star kicks than CCS, but may be more constraining than other
estimates.

With a firm estimate of the efficiency for converting WDs to NRPs and
the duration of this process, a more exact estimate could be made.
There are two important caveats to keep in mind, however.  The first
is that model classes L and G are based on the upper limit of the
number of NRPs in only one cluster, M22.  We chose this cluster
because it has been searched to a lower limiting luminosity than any
other.  If M22 is an atypical cluster in terms of its NRP content,
then this would bias our results.  However, other clusters (e.g. M28,
47~Tucanae, Terzan 5) have been deeply searched and also have small
implied NRP birth rates.  Model classes L and G also assume that only
one characteristic of a GC influences the number of NRPs that are
formed.  \citetalias{blt+11} found evidence suggesting that
metallicity may play a significant role in the formation of NRPs, and
we cannot rule out other factors.

The best way to differentiate between these results and those
presented in \citetalias{blt+11} would be to search a large number of
GCs more deeply.  This would yield much tighter constraints for the
method used in \citetalias{blt+11}, and would reduce the number of
simplifying assumptions we need to make here.

\subsection{Can the Population of Escaped NRPs Be Identified?
  \label{sec:pop_detect}}

The next question to ask is, what are the prospects for identifying
the population of escaped NRPs as separate from normal disk pulsars?
In the following discussion, we will limit ourselves to the results
for model class L.  With large numbers of pulsars it may be able to
identify NRPs based on their spatial distribution around GCs or their
kinematic similarities, since NRPs born from the same GC should have
the same systemic contributions to their velocities.  Unfortunately,
we favor models with low $\sigma_v$, and as \S\ref{sec:surv_results}
makes clear, in this case even the best surveys can detect $\lesssim
100$ NRPs.  Proper motions would probably only be available on a
sub-set of these pulsars.  However, when the SKA comes online, it will
be able to measure proper motions interferometricaly on nearly all
pulsars it detects, so this may be a future avenue for obtaining
proper motions.  Nonetheless, it is worth keeping our results in mind
as high sensitivity surveys are carried out.

Large area surveys are not the only means by which GC NRPs could be
detected.  Since many escaped NRPs are found to remain fairly close to
their progenitor clusters, especially for our favored low $\sigma_v$
models, targeted surveys of the regions around GCs could be more
fruitful, as it would allow for deeper integrations and better
sensitivity.  We have calculated the minimum distance between an
escaped NRP and a GC for all emitting pulsars in our simulation.
There is some dependence on $\sigma_v$, but in general $\sim
3$--$7\%$, $\sim 13$--$16\%$, and $\sim 20$--$25\%$ of NRPs lie within
$1^{\circ}$, $3^{\circ}$, and $5^{\circ}$ of a GC, respectively.
However, a large $\sigma_v$ would still give rise to a larger pool of
potentially observable pulsars.  A survey covering a $\sim 3^{\circ}$
radius around all GCs would then, in principle, capture roughly $15\%$
of the population of NRPs.  However, with 156 Milky Way GCs, this
would amounts to a total survey area of over 4400 square degrees, with
each portion requiring substantial integration times.  Such a survey
would be very difficult.  The situation is improved if such a survey
were limited to only $1^{\circ}$ around each cluster (reducing the
survey area to about 500 square degrees) or by targeting only the most
promising GCs, such as those already known to contain an NRP, but this
would also decrease the likely number of NRPs that could be detected.

\section{Conclusions \label{sec:conc}}

\citetalias{blt+11} explored the enigmatic population of NRPs found in
GCs by setting limits on the size of the population using the results
of various GC surveys.  We have built upon these results to explore
the properties of NRPs that escape from clusters and enter the field
of the Galaxy.  We agree with \citetalias{blt+11} that current GC
surveys are not sensitive enough to constrain the population on their
own.  However, if we assume that the number of NRPs in a GC scale with
some properties of the cluster, specifically luminosity (as a proxy
for cluster mass), then the size of the population is more realistic.
It seems that in these cases, ECS may be sufficient to account for GC
NRPs.  We favor models that rely on luminosity scaling and invoke a
low natal kick velocity ($\sim 10\; \km\, \ps$), but there is still
too much uncertainty in how NRPs are formed to place more definite
limits.  Unfortunately, large numbers of escaped NRPs would be
difficult to detect with any current large-area pulsar surveys, so the
chances of identifying them as a separate population are slim.  The
best prospects lie with future surveys with the SKA.  If sufficiently
large numbers of GC NRPs \emph{can} be detected in the field of the
Galaxy, it may be possible to distinguish them from normal disk
pulsars through their spatial distribution and proper motions.  The
best approach for learning more about NRPs, though, remains
identifying those that are bound to their progenitor GCs.  This will
require more sensitive searches of many clusters.

We are extremely grateful to Norbert Wex for providing us with an
independent method for checking the results of our simulations.  We
thank an anonymous referee for reviewing this manuscript.  We would
also like to thank the National Science Foundation for supporting this
work through grant AST-0907967.  The National Radio Astronomy
Observatory is a facility of the National Science Foundation operated
under cooperative agreement by Associated Universities, Inc.

\bibliographystyle{apj} 
\bibliography{ms}{}

\begin{deluxetable}{ccccc}
  \centering 
  \tabletypesize{\normalsize} 
  \tablewidth{0pt} 
  \tablecolumns{5}
  \tablecaption{Properties of M22 Used for Scaling \label{table:M22}}
  \tablehead{
    \colhead{$N_{95}$}              &
    \colhead{$v\rmsub{esc}$}        &
    \colhead{$L\rmsub{V}$}          &
    \colhead{$\log{\rho\rmsub{c}}$} & 
    \colhead{$r\rmsub{c}$}          \\ 
    \colhead{}                      &
    \colhead{($\km\, \ps$)}         &
    \colhead{($L\rmsub{\Sun,V}$)}   &
    \colhead{($\Lsun\, \pcubpc)$}   &
    \colhead{(\pc)}}                
  \startdata
    4 & 44.7 & $2.15 \times 10^{5}$ & 3.63 & 1.2 \\
  \enddata
  \tablecomments{Structural parameters have been taken from
    \citet{har10}, and we have used $D = 3.2\; \kpc$ for calculating
    $r_c$ in physical units.  The escape velocity is from
    \citet{gzp+02}.}
\end{deluxetable}

\begin{deluxetable}{ccc}
  \centering 
  \tabletypesize{\normalsize} 
  \tablewidth{0pt} 
  \tablecolumns{3}
  \tablecaption{Summary of Models Used to Simulate the Population of
    NRPs \label{table:models}}
  \tablehead{
    \colhead{Model Class}   &
    \colhead{$N\rmsub{GC}$} &
    \colhead{Scaling}}
  \startdata
  Sa\tablenotemark{a} & 97  & None              \\
  Sb\tablenotemark{b} & 81  & None              \\
  L\tablenotemark{c}  & 156 & Luminosity        \\
  G\tablenotemark{c}  & 143 & $\Gamma\rmsub{c}$ \\
  \enddata 
  \tablecomments{Ten runs were performed for each value of $\sigma_v$
    for each model class.}
  \tablenotetext{a}{Sa models use all the GCs studied in
    \citetalias{blt+11}.}
  \tablenotetext{b}{Sb models exclude GCs with $L\rmsub{1.4\; \GHz,
      min} \geq 7.5\; \mJy\, \kpc^2$ because these shallow surveys
    were not very constraining.}
  \tablenotetext{c}{M22 was used as the reference cluster when
    scaling.}
\end{deluxetable}

\begin{deluxetable}{lccccccccc}
  \centering
  \tabletypesize{\footnotesize}
  \tablewidth{0pt}
  \tablecolumns{10}
  \tablecaption{Parameters Used in the Model of Galactic Potential
    \label{table:phi_params}}
  \tablehead{
    \colhead{Component}        &
    \colhead{$M$}              &
    \colhead{$a$}              &
    \colhead{$b$}              &
    \colhead{$\beta_1$}        &
    \colhead{$h_1$}            &
    \colhead{$\beta_2$}        &
    \colhead{$h_2$}            &
    \colhead{$\beta_3$}        &
    \colhead{$h_3$}            \\
    \colhead{}                 &
    \colhead{($10^9\; \Msun$)} &
    \colhead{(\kpc)}           &
    \colhead{(\kpc)}           &
    \colhead{}                 &
    \colhead{(\kpc)}           &
    \colhead{}                 &
    \colhead{(\kpc)}           &
    \colhead{}                 &
    \colhead{(\kpc)}}
  \startdata
    Disk-Halo  & 145 & 2.4 & 5.5  & 0.4 & 0.325 & 0.5 & 0.090 & 0.1 & 0.125 \\
    Bulge      & 9.3 & 0   & 0.25 & 1   & 0     & 0   & 0     & 0   & 0     \\
    Nucleus    & 10  & 0   & 1.5  & 1   & 0     & 0   & 0     & 0   & 0     \\
  \enddata
  \tablecomments{See \S\ref{sec:xvevol} for parameter definitions.}
\end{deluxetable}

\begin{deluxetable}{lccccccccc}
  \centering 
  \tabletypesize{\footnotesize} 
  \tablewidth{0pt} 
  \tablecolumns{10}
  \tablecaption{Large-area Survey Parameters \label{table:surveys}}
  \tablehead{
    \colhead{Survey}           &
    \colhead{$f\rmsub{sky}$\tablenotemark{a}}   &
    \colhead{$G$}              &
    \colhead{$T\rmsub{sys}$}   &
    \colhead{$\nu\rmsub{obs}$} &
    \colhead{$\Delta \nu$}     &
    \colhead{$t\rmsub{samp}$}  &
    \colhead{$t\rmsub{int}$}   &
    \colhead{$\beta$}          &
    \colhead{$S\rmsub{1374\; MHz, min}$\tablenotemark{b}} \\
    \colhead{}                 &
    \colhead{(\%)}             &
    \colhead{($\kelvin\; \Jy^{-1}$)} &
    \colhead{(\kelvin)}        &
    \colhead{(\MHz)}           &
    \colhead{(\MHz)}           &
    \colhead{($\us$)}          &
    \colhead{(\s)}             &
    \colhead{}                 &
    \colhead{(\uJy)}}
  \startdata
    PMSURV & 4.6     & 0.6 & 25 & 1374 & 288 & 250   & 2100 & 1.2 & 160  \\
    P-ALFA & 10.9    & 8.5 & 25 & 1374 & 300 & 64    & 268  & 1.1 & 100  \\
    GBNCC  & 19.2    & 2.0 & 46 & 350  & 100 & 81.92 & 120  & 1.1 & 2000 \\
    GBTALL & 86.0    & 2.0 & 46 & 350  & 100 & 81.92 & 120  & 1.1 & 2000 \\
    SKA    & 100     & 140 & 25 & 1374 & 512 & 50    & 2100 & 1.0 & 0.45 \\
  \enddata
  \tablecomments{All surveys use two summed polarizations.}
  \tablenotetext{a}{Fractional sky coverage}
  \tablenotetext{b}{Approximte limiting flux density scaled to
    $1374\; \MHz$ assuming a spectral index of $-1.6$}
\end{deluxetable}

\begin{deluxetable}{lcccccccccc}
  \centering 
  \tabletypesize{\scriptsize} 
  \tablewidth{0pt} 
  \tablecolumns{9}
  \tablecaption{Simulation Results \label{table:results}}
  \tablehead{
    \colhead{Model ID\tablenotemark{a}}       &
    \colhead{$N\rmsub{esc}$\tablenotemark{b}} &
    \colhead{$N\rmsub{ret}$\tablenotemark{b}} &
    \colhead{$N\rmsub{emit}$}                 &
    \multicolumn{5}{c}{$N\rmsub{det}$\tablenotemark{c}}       \\
    \colhead{}                                &
    \colhead{}                                &
    \colhead{}                                &
    \colhead{}                                &
    \colhead{PMSURV}                          &
    \colhead{PALAFA}                          &
    \colhead{GBNCC}                           &
    \colhead{GBTALL}                          &
    \colhead{SKA}}
  \startdata
    Sa5   & $1.3 \times 10^6$ & $9.5 \times 10^5$ & $1.5 \times 10^5$ & $3$               & $0$               & $0$               & $83$              & $1.1 \times 10^4$ \\
    Sa10  & $1.0 \times 10^7$ & $9.5 \times 10^5$ & $1.2 \times 10^6$ & $40$              & $6$               & $0$               & $2.6 \times 10^2$ & $9.2 \times 10^4$ \\
    Sa20  & $8.0 \times 10^7$ & $9.7 \times 10^5$ & $9.1 \times 10^6$ & $3.9 \times 10^2$ & $51$              & $0$               & $3.5 \times 10^3$ & $6.9 \times 10^5$ \\
    Sa30  & $2.7 \times 10^8$ & $1.0 \times 10^6$ & $3.1 \times 10^7$ & $3.0 \times 10^2$ & $7.0 \times 10^2$ & $0$               & $7.9 \times 10^3$ & $2.4 \times 10^6$ \\
    Sa50  & $1.2 \times 10^9$ & $9.2 \times 10^5$ & $1.4 \times 10^8$ & $5.4 \times 10^3$ & $5.9 \times 10^2$ & $0$               & $5.4 \times 10^4$ & $1.1 \times 10^7$ \\
    Sa70  & $3.4 \times 10^9$ & $9.7 \times 10^5$ & $4.2 \times 10^8$ & $1.0 \times 10^4$ & $2.4 \times 10^3$ & $1$               & $1.1 \times 10^5$ & $3.0 \times 10^7$ \\
    Sa100 & $9.9 \times 10^9$ & $9.3 \times 10^5$ & $1.1 \times 10^9$ & $6.1 \times 10^4$ & $5.4 \times 10^3$ & $35$              & $3.7 \times 10^5$ & $8.3 \times 10^7$ \\
    \hline
    Sb5   & $3.4 \times 10^4$ & $4.7 \times 10^5$ & $3.5 \times 10^3$ & $0$               & $0$               & $0$               & $2$               & $3.8 \times 10^2$ \\
    Sb10  & $3.8 \times 10^5$ & $4.8 \times 10^5$ & $3.6 \times 10^4$ & $6$               & $6$               & $0$               & $23$              & $4.3 \times 10^3$ \\
    Sb20  & $3.3 \times 10^6$ & $4.9 \times 10^5$ & $3.3 \times 10^5$ & $63$              & $62$              & $0$               & $2.1 \times 10^2$ & $3.8 \times 10^4$ \\
    Sb30  & $1.1 \times 10^7$ & $5.1 \times 10^5$ & $1.1 \times 10^6$ & $2.5 \times 10^2$ & $1.9 \times 10^2$ & $0$               & $5.0 \times 10^2$ & $1.3 \times 10^5$ \\
    Sb50  & $5.2 \times 10^7$ & $5.2 \times 10^5$ & $4.9 \times 10^6$ & $9.0 \times 10^2$ & $7.2 \times 10^2$ & $0$               & $2.4 \times 10^3$ & $5.8 \times 10^5$ \\
    Sb70  & $1.4 \times 10^8$ & $5.6 \times 10^5$ & $1.3 \times 10^7$ & $2.6 \times 10^3$ & $2.8 \times 10^3$ & $2$               & $7.0 \times 10^3$ & $1.6 \times 10^6$ \\
    Sb100 & $4.1 \times 10^8$ & $5.1 \times 10^5$ & $4.1 \times 10^7$ & $7.6 \times 10^3$ & $6.5 \times 10^3$ & $40$              & $2.8 \times 10^4$ & $4.4 \times 10^6$ \\
    \hline
    L5    & $2.0 \times 10^2$ & $1.5 \times 10^4$ & $42$              & $0$               & $0$               & $0$               & $0$               & $2$               \\
    L10   & $8.8 \times 10^2$ & $1.4 \times 10^4$ & $1.7 \times 10^2$ & $0$               & $0$               & $0$               & $0$               & $10$              \\
    L20   & $5.0 \times 10^3$ & $1.3 \times 10^4$ & $9.7 \times 10^2$ & $0$               & $0.17$            & $0$               & $0$               & $58$              \\
    L30   & $1.6 \times 10^4$ & $1.5 \times 10^4$ & $3.2 \times 10^3$ & $1$               & $0$               & $0$               & $1$               & $1.9 \times 10^2$ \\
    L50   & $6.9 \times 10^4$ & $2.0 \times 10^4$ & $1.4 \times 10^4$ & $5$               & $1$               & $0$               & $6$               & $9.4 \times 10^2$ \\
    L70   & $1.9 \times 10^5$ & $2.7 \times 10^4$ & $3.8 \times 10^4$ & $14$              & $3$               & $0$               & $16$              & $2.7 \times 10^3$ \\
    L100  & $5.9 \times 10^5$ & $3.4 \times 10^4$ & $1.2 \times 10^5$ & $38$              & $9$               & $0.57$            & $47$              & $7.8 \times 10^3$ \\
    \hline
    G5    & $53$              & $7.3 \times 10^4$ & $7$               & $0$               & $0$               & $0$               & $0$               & $1$               \\
    G10   & $1.1 \times 10^3$ & $7.2 \times 10^4$ & $2.0 \times 10^2$ & $0.11$            & $0$               & $0$               & $0$               & $15$              \\
    G20   & $1.7 \times 10^4$ & $7.0 \times 10^4$ & $3.3 \times 10^3$ & $3$               & $0.52$            & $0$               & $1$               & $2.5 \times 10^2$ \\
    G30   & $6.0 \times 10^4$ & $9.7 \times 10^4$ & $1.1 \times 10^4$ & $9$               & $1$               & $0$               & $3$               & $8.5 \times 10^2$ \\
    G50   & $3.2 \times 10^5$ & $1.9 \times 10^5$ & $6.0 \times 10^4$ & $49$              & $4$               & $0$               & $19$              & $4.6 \times 10^3$ \\
    G70   & $1.0 \times 10^6$ & $2.6 \times 10^5$ & $1.9 \times 10^5$ & $1.6 \times 10^2$ & $8$               & $0$               & $58$              & $1.4 \times 10^4$ \\
    G100  & $3.3 \times 10^6$ & $3.2 \times 10^5$ & $6.0 \times 10^5$ & $4.6 \times 10^2$ & $24$              & $1$               & $2.3 \times 10^2$ & $4.3 \times 10^4$ \\
  \enddata
  \tablenotetext{a}{Model IDs include model class and $\sigma_v$ (e.g.
    Sa5 refers to model class Sa and $\sigma_v = 5$).}
  \tablenotetext{b}{$N\rmsub{esc}$ and $N\rmsub{ret}$ are all escaped
    and retained pulsars produced in the last $200\; \Myr$.  These
    numbers scale linearly with the chosen timescale.}
  \tablenotetext{c}{$N\rmsub{det}$ has already been corrected for a
    10\% beaming fraction.}
  \tablecomments{The relative success of some surveys (such as the
    PMSURV, PALFA, and GBTALL) depends slightly on model class.  This
    is because some model classes exclude certain clusters.  Pulsars
    from these clusters will be concentrated in a certain region of
    the sky, but surveys do not have identical sky coverage.}
\end{deluxetable}

\begin{deluxetable}{lcc}
  \centering 
  \tabletypesize{\footnotesize} 
  \tablewidth{0pt} 
  \tablecolumns{3}
  \tablecaption{Birth Rates of GC NRPs \label{table:birthrates}}
  \tablehead{
    \colhead{Model ID}                        &
    \colhead{$\mathcal{R}$}                   &
    \colhead{$\tau\rmsub{exhaust}$\tablenotemark{a}} \\
    \colhead{}                                &
    \colhead{($\mathrm{psr\, century^{-1}}$)} &
    \colhead{($\Myr$)}}
  \startdata
    Sa5    & $1.1$    & $19$     \\
    Sa10   & $5.5$    & $3.8$    \\
    Sa20   & $40$     & $0.52$   \\
    Sa30   & $140$    & $0.15$   \\
    Sa50   & $600$    & $0.035$  \\
    Sa70   & $1700$   & $0.012$  \\
    Sa100  & $5000$   & $0.0042$ \\
    \hline
    Sb5    & $0.25$   & $83$     \\
    Sb10   & $0.43$   & $49$     \\
    Sb20   & $1.9$    & $11$     \\
    Sb30   & $5.8$    & $3.6$    \\
    Sb50   & $26$     & $0.8$    \\
    Sb70   & $70$     & $0.3$    \\
    Sb100  & $210$    & $0.1$    \\
    \hline
    L5     & $0.0076$ & $2800$   \\
    L10    & $0.0074$ & $2800$   \\
    L20    & $0.009$  & $2300$   \\
    L30    & $0.016$  & $1400$   \\
    L50    & $0.044$  & $470$    \\
    L70    & $0.11$   & $190$    \\
    L100   & $0.31$   & $67$     \\
    \hline
    G5     & $0.037$  & $570$    \\
    G10    & $0.037$  & $570$    \\
    G20    & $0.044$  & $480$    \\
    G30    & $0.079$  & $270$    \\
    G50    & $0.26$   & $82$     \\
    G70    & $0.63$   & $33$     \\
    G100   & $1.8$    & $12$     \\
  \enddata
  \tablenotetext{a}{The time scale to exhaust a supply of $\sim 2.4
    \times 10^5$ WDs via ECS, assuming a constant birthrate (see
    text).}
\end{deluxetable}

\end{document}